\documentclass{ws-mpla}

\usepackage{subfigure}

\def\beq{\begin{equation}}
\def\eeq{\end{equation}}
\def\bea{\begin{eqnarray}}
\def\eea{\end{eqnarray}}

\def\a{\alpha}

\def\P{\Pi}
\def\r{\rho}    
\def\p{\pi}
\def\cd{{\cal D}}
\def\d{\delta}
\def\g{\gamma}
\def\b{\beta}

\def\m{\mu}
\def\n{\nu}
\def\MSbar{$\overline{\rm MS}$}

\def\nn{\nonumber}

\begin{document}

\markboth{D. Boito, M. Golterman, K. Maltman, S. Peris}
{$\alpha_s$ analyses from hadronic tau decays with OPAL and ALEPH data }


\title{$\alpha_s$ ANALYSES FROM HADRONIC TAU DECAYS WITH OPAL AND ALEPH DATA
}

\author{DIOGO BOITO\footnote{Speaker}}
\address{S\~ao Carlos Institute of Physics, University of S\~ao Paulo\\
PO Box 369, 13570-970, S\~ao Carlos, SP, Brazil\\
boito@ifsc.usp.br}

\author{MAARTEN GOLTERMAN}
\address{Department of Physics and Astronomy, San Francisco State University\\
San Francisco, CA 94132, USA
}

\author{KIM MALTMAN}
\address{Department of Mathematics and Statistics, York University,\\
Toronto, ON Canada M3J 1P3,\\  and \\CSSM, University of Adelaide, \\Adelaide, SA 5005 Australia}

\author{SANTIAGO PERIS}
\address{Department of Physics, Universitat Aut\`onoma de Barcelona\\
E-08193 Bellaterra, Barcelona, Spain}

\maketitle


\begin{abstract}

Recently, we extracted the strong coupling, $\alpha_s(m_\tau^2)$, from the revised
ALEPH data for non-strange hadronic tau decays. Our analysis is based
on a method previously used for the determination of the
strong coupling from OPAL data.  In our strategy, we employ different moments of the
spectral functions
both with and without pinching, including Duality Violations, in order
to obtain fully self-consistent
analyses that do not rely on untested assumptions (such as the
smallness of higher dimension contributions in the OPE).
Here we  discuss the $\alpha_s$ values obtained from the ALEPH and the
OPAL data, the robustness of the analysis, as well as non-perturbative contributions  from DVs and the OPE. We show that, although the $\alpha_s$
determination is sound, non-perturbative effects limit the accuracy
with which one can extract the strong coupling from tau decay data.
Finally, we discuss the compatibility of the data sets and the possibility
of a combined analysis.
\end{abstract}


\section{Introduction}

Since the 90s, hadronic tau decays have been used to extract the QCD
coupling, $\alpha_s$. One of the appealing features of this
determination is that it represents a non-trivial test of the
$\alpha_s$ evolution as predicted by the celebrated QCD
$\beta$-function. The scale set by the $\tau$ mass is rather low,
$m_\tau \approx 1.78$~GeV, but it still allows for a meaningful
perturbative treatment, provided also non-perturbative contributions
are taken into account. 
It has become standard to organize the QCD description of these decays
in the form of Wilson's operator product expansion (OPE).\cite{BNP92} In this
OPE, apart from the purely perturbative contribution and quark-mass
corrections, non-perturbative QCD condensates also occur.

The extraction of $\alpha_s$ is performed through the use of finite
energy sum rules (FESR). Observables such as the ratio 
\beq\label{RtauDef}
R_\tau =\frac{\Gamma \left[ \tau^- \to \nu_\tau {\rm hadrons}(\gamma)  \right]}{\Gamma \left[ \tau^- \to \nu_\tau e^-\bar \nu_e(\gamma)  \right]},
\eeq
can be written as weighted integrals over the experimentally
accessible QCD spectral functions, that can be reconstructed from the
measurement of the dominant exclusive channels in the decays $\tau \to
({\rm hadrons}) + \nu_\tau$.\cite{Tsai71}  The integrals of experimental data
are performed over the total hadronic momentum $s$ and run from zero
 to $m_\tau^2$. Clearly, the OPE description is not valid in the low energy part
of this interval. One then resorts to the analyticity properties of the
QCD correlators to write the theoretical counterpart of the weighted spectral integral as
an integral along a closed circle of radius $|s|=m_\tau^2$ in the
complex plane. Additional sum rules, apart from the one giving $R_\tau$,
can be constructed by using different  weight functions. This
freedom is exploited in order to constrain additional parameters of the
OPE, such as QCD condensates, and extract them in combination with $\alpha_s$.

The QCD spectral functions (in the vector and axial-vector channels)
were determined originally from hadronic tau decays by the LEP
collaborations ALEPH and OPAL, in the 90s.\cite{ALEPH,OPAL} Recently,
a re-analysis of the ALEPH data was published.\cite{Davieretal14} This
analysis was performed with a different binning and employing a new
unfolding method. The new analysis corrects a problem in the
older version of the correlation matrices.\cite{Manchester} Since this
correction, ALEPH's can be considered as the best data set, since it
has smaller uncertainties.

 On the theory side, two aspects of these $\alpha_s$ determinations
 have received special attention, recently. The first regards the use
 of the Renormalisation Group in the improvement of the perturbative
 series.  Several prescriptions are advocated in the literature, among
 which two stand out the most: Contour Improved Perturbation Theory
 (CIPT)\cite{CIPT,CIPT2} and Fixed Order Perturbation Theory
 (FOPT).\cite{FOPT} The two prescriptions lead to different
 perturbative series and, hence, to different values of $\alpha_s$
 when used in $\alpha_s$ extractions. The differences have not
 diminished with the computation of the NNNLO term,
 $\mathcal{O}(\alpha_s^4)$, in the perturbative expansion.\cite{BCK08}
 This discrepancy remains one of the main sources of theoretical error
 associated with $\alpha_s$ from $\tau$ decays. Arguments in favour of
 FOPT have been put forward recently,\cite{BJ08,BBJ13,MJMainz} but the
issue is still being debated.\cite{CF11,AACF13} We
 prefer to remain conservative and always quote two values of
 $\alpha_s$ from our analyses.

The second aspect that has received attention in the past few years is
related to the non-perturbative contributions. Since the work of
Ref.~\refcite{MY08}, it has become clear that the treatment of the
higher order condensates advocated in some of the recent $\alpha_s$
analyses is inconsistent.  The OPE alone, moreover, cannot account for
all non-perturbative effects in the vicinity of the Minkowski axis;
duality violations (DV) are present and should be included in the
theoretical description. However, since the kinematic weight function
related to $R_\tau$ possesses a  double zero on the positive axis at $s=m_\tau^2$
that suppresses contributions from this region, the DV part of the
correlators was often disregarded. For this reason, older analyses
were restricted to the so-called {\it pinched moments} --- those
moments that also exhibit a zero on the axis. The problem with that
strategy is that more pinching enhances higher-dimension contributions
in the OPE, which augments the number of  parameters to be
determined from the experimental data. One way around this complication,
pursued in a number of analyses, was to simply assume that contributions
of dimension higher than 8 could be neglected.
However, it was shown in Ref.~\refcite{MY08} that this leads to
results that do not survive self consistency checks --- they 
provide poor matches to the corresponding spectral integrals when $s_0$ is lowered below $m_\tau^2$.  In conclusion, the assumption is too strong and results
based on it carry an unquantified systematic uncertainty.\footnote{We
  refer to S. Peris' contribution to these proceedings for a more detailed
  discussion about this issue.\cite{PerisMainz} See also in
  Section~VII of Ref.~\refcite{alphas2014}.}

Recently, thanks to progress in modelling the DV
contributions,\cite{BlokShifmanetal,CGP05,CGP08,CGP09} it has become
possible to include them in a fully self-consistent analysis of
$\alpha_s$.\cite{alphas2011,alphas2012,alphas2014} In this new
framework, no additional assumptions regarding the OPE are required: at
each order in the OPE, the leading contribution is taken into account
and, as a consequence, we find that the results thus obtained pass all
consistency checks.  After a brief recollection of the theoretical
framework in Sec. 2, we discuss, in Sec. 3, results from analyses
following this new strategy obtained from ALEPH and OPAL spectral
function data. An issue that is still open is to what extent one can
combine results obtained from the two data sets, and whether or not a
combined analysis is justified. This question will be touched on in
Sec.~\ref{Comparison}. In Sec.~\ref{Conclusions} we present our
conclusions.

\section{Analysis framework}

The purpose of this section is  to provide a brief review of the
framework of our analysis. The details can be found in the original
publications, Refs.~\refcite{alphas2011},~\refcite{alphas2012} and \refcite{alphas2014}.

We employ FESRs of the following  form\cite{BNP92,Shankaretal}
\bea
\label{FESR}
I^{(w)}_{V/A}(s_0)&\equiv&\int_0^{s_0}\frac{ds}{s_0}\;w(s)\;\r^{(1+0)}_{V/A}(s)
=-\frac{1}{2\p i}\oint_{|s|=s_0}\frac{ds}{s_0}\;w(s)\;\P^{(1+0)}_{V/A}(s)\ ,
\eea
where $\rho^{(1+0)}_{V/A}$ is the
experimentally accessible spectral function and  the weight-functions $w(s)$ are  polynomials in $s$. The correlators 
$\P^{(1+0)}_{V/A}(s)$ are given by
\begin{align}
\label{corr}
&i\int d^4x\,e^{iqx}\,\langle 0|T\left\{J_\m(x)J_\n^\dagger(0)\right\}|0\rangle
\nn\\ &=\left(q_\m q_\n-q^2 g_{\m\n}\right)\P^{(1+0)}(s)+q^2 g_{\m\n}\P^{(0)}(s)\ ,
\end{align}
with $s=q^2=-Q^2$. The current $J_\m$ is one of the non-strange $V$ or
$A$ currents, namely, $\bar u\g_\m d$ or $\bar u\g_\m\g_5 d$. The
superscripts $(0)$ and $(1)$ refer to spin.  For a given weight
function, we construct FESRs at several values of $s_0\leq m_\tau^2$ .

The correlators  $\P^{(1+0)}_{V/A}(s)$ admit a decomposition into
three  parts
\begin{equation}
\label{th}
\P^{(1+0)}(s)=\P^{(1+0)}_{ \rm pert}(s)+\P^{(1+0)}_{\rm OPE}(s)+\P^{(1+0)}_{\rm DV}(s).
\end{equation}
In the above, ``pert'' denotes perturbative (which can be regarded as
the dimension zero contribution to the OPE), ``OPE'' refers to OPE
corrections of dimension larger than zero (including quark-mass
corrections), whereas ``DV'' denotes the DV contributions to
$\P^{(1+0)}(s)$.

The ambiguity related to the use of the renormalisation group
to which we alluded  affects (mainly) the perturbative part
of the correlators. When treating the contour
integration in the FESR, one must adopt a prescription for the renormalisation
scale.  As discussed above, we perform our analysis using CIPT or
FOPT and quote results for both on an equal footing. 

Contributions from higher dimensions in the OPE 
are parametrised  with effective condensates $C_{D}$ as
\begin{equation}
\label{OPE}
\P^{(1+0)}_{\rm OPE}(s)=\sum_{k=1}^\infty\frac{C_{2k}(s)}{(-s)^k}\ .
\end{equation}
 Since we work with non-strange spectral functions, the dimension-two
 quark-mass corrections can safely be neglected.\footnote{This has
   been checked explicitly.}  Therefore, in our analysis, the
 dimension two contribution is absent. In principle, the first
 non-negligible contribution is dimension 4, encoded in $C_4$, that
 can be related to the gluon and quark condensates.  However, the weight
 functions employed in our analysis (see below) are polynomials
 constructed from combinations of  unity, $s^2$, and $s^3$. As a
 result, in our FESRs, the leading contributions from the OPE arise
 solely from $C_6$ and $C_8$.  Subleading logarithmic corrections to
 these coefficients are neglected and all $C_D$ are treated as
 constants.

Through the use of analyticity, the DV contribution to the sum rules
can be cast as\cite{CGP09}  
\begin{equation}
\label{FESRDV}
\cd_w(s_0)=-\int_{s_0}^\infty \frac{ds}{s_0}\,w(s)\,\r^{\rm DV}(s),
\end{equation}
where $\r^{\rm DV}(s)$ is the DV part of the spectral function in a given channel \beq
\label{specDV}
\r^{\rm DV}(s)=\frac{1}{\p}\,\mbox{Im}\,\P^{(1+0)}_{\rm DV}(s).
\end{equation}
For $s$ large enough,  the DV part of the spectral function can be parametrised with the Ansatz of Refs.~\refcite{CGP05,CGP08}
\begin{equation}
\label{DVpar}
\r^{\rm DV}_{V/A}(s)=\mbox{exp}\left(-\d_{V/A}-\g_{V/A}s\right)\sin\left(\a_{V/A}+\b_{V/A}s\right) .
\end{equation}
The parameters associated with DV are, in principle, channel
dependent. There is no a priori reason to treat them as equal in $V$
and $A$ since they depend, ultimately, on the large-$s$ behaviour of the
QCD resonances that appear in each channel. Therefore, we keep them
different in $V$ and $A$, thus avoiding  any additional assumptions.  The parametrisation
of Eq.~(\ref{DVpar})
adds four new parameters per channel.

Since we include the DV contribution, our analysis does not have to be
restricted to pinched weight functions. Rather, a weight
function without pinching is included in order to constrain the DV parameters
better. After  extensive explorations in Refs.~\refcite{alphas2011}
and~\refcite{alphas2012} we concluded that it is sufficient to work
with three weight functions, namely,
\bea
w_1&=&1, \nn\\ 
w_2&=&1-x^2, \nn \\
w_3&=& (1-x)^2(1+2x), 
\label{moments}
\eea
where $x\equiv s/s_0$. The weight function $w_3$ appears
multiplying the $J=0+1$ spectral function in the expression of
$R_\tau$,  of
Eq.~(\ref{RtauDef}),  and is hence often called the  ``kinematical weight.''
  The main motivation behind this choice
is the fact that we want to perform a self-consistent
analysis, including all leading order contributions in the OPE,
without making any untested assumption about higher order
condensates. The  explorations of Refs.~\refcite{alphas2011}
and~\refcite{alphas2012}  established that this set of weight
functions fulfils these requirements and allows for a good
determination of $\alpha_s$. As a final remark,
we observe that these three weight
functions  have good perturbative behaviour, in the sense of the
analysis performed in Ref.~\refcite{BBJ13}.

\section{Fits and main results}
\subsection{Fits}

In our analyses we have performed several different fits: to
individual channels, $V$ or $A$, or combining $V$ and $A$ together; fits
with a single weight function or combining different subsets of the
weight functions given in Eq.~(\ref{moments}). (In the combined
$V$ and $A$ fits the equality of $\alpha_s$ in the two channels is, of course, always
imposed.)  Also, we always employ
a window of $s_0$  values $[s_{\rm min}, s_{\rm max}]$ and we make
sure to test the stability of the results under variations of this
window. The window must be chosen such as to maintain  the validity of the
description of the QCD correlator through Eq.~(\ref{th}).

Fits that involve a single weight function are performed minimising a
standard $\chi^2$, including all correlations among the different
moments. Simultaneous fits to more than one weight function, on the
other hand, are too strongly correlated to allow for the use of a fit
quality of this type. Alternative fit qualities must be employed in
these cases. The error  propagation (including all correlations)
is done according to a standard procedure, described
in detail in the appendix of Ref.~\refcite{alphas2011}.

Several consistency checks are performed on the fit results to assess
their robustness. On the statistical side, we have performed a study of
the posterior probability in the parameter space using Markov-chain
Monte Carlo simulations. This was particularly helpful in the case of
OPAL data, since this data set has larger uncertainties that produce a
shallower fit quality, which oftentimes displays multiple minima.  On
the physical side, the results must be as immune as possible to
changes in the $s_0$ window employed in the fits. To maximize the use
of the data, we always integrated the experimental results up to the
last bin, $s_{\rm max} = m_\tau^2$. The lower edge of the $s_0$
window, $s_{\rm min}$, was varied between 1.3 and 1.7~GeV$^2$ to check
for the stability of the results.\footnote{The exact values used in
  this variation depend on the data set, since the binning is
  different.} Other important tests that  we implemented are the
Weinberg sum rules.\footnote{We checked that our results fulfil both the first and second Weinberg sum rules\cite{WSR} as well as the sum rule for the pion
electromagnetic mass splitting.\cite{DGMLY}} The results of our fits including both $V$ and $A$
channels fulfil these sum rules within the whole fit window
employed in our analyses.

\subsection{Results for $\alpha_s$ from the OPAL- and ALEPH-based analyses}

 Our fits
determine simultaneously $\alpha_s$, and non-perturbative contributions, such as  OPE condensates as well as
the DV parameters that were introduced in Eq.~(\ref{DVpar}). Here, we choose to focus mainly on
$\alpha_s$ results. The
results are given in full detail in the original publications.\cite{alphas2012,alphas2014}

At first, we restricted our attention to OPAL data.\cite{alphas2011}
The reason why we focussed, at the time, on OPAL's data is the fact
that it was discovered that the correlation matrix of the then
publicly available ALEPH spectral functions missed a contribution
from the unfolding procedure.\cite{Manchester} Therefore, we decided to perform
an update of the original OPAL spectral functions to reflect modern
values of branching fractions and constants used for normalisation.\footnote{The updated
  version of the OPAL spectral functions can be provided upon
  request.} The results described here are based on the analysis
of these updated OPAL spectral functions, first reported in Ref.~\refcite{alphas2012}.

The $\alpha_s$
values that we obtained from the different fit set-ups described above
are consistent within their somewhat large error bars (dominated by
statistical errors of the OPAL data). However, fits including the $A$
channel do require an additional assumption, related to the
larger mass of the $a_1$ resonance, as compared to the $\rho$. Since
the parametrisation of DV  that we employed is based on the
asymptotic regime of the QCD resonances, one must assume that this
regime is already reached close to the tail of the $a_1$.
The overall consistency of our results seems to indicate that this
assumptions is fulfilled within uncertainties. However, to avoid
having an unquantified systematic  from this extra assumption, we prefer
to quote as final values those that arise from fits to the $V$ channel
only.

In the case of the analysis based on OPAL data, our final values come
from the analysis of the $V$ channel with the weight function
$w_1=1$. This choice is motivated by the wider range of stability of
these results against variations of $s_{\rm min}$ although, again, all our
results are consistent among themselves. In the \MSbar\  scheme and with $N_f=3$
we found
\begin{align}
& \alpha_s(m_\tau^2) = 0.325 \pm 0.018 \qquad ({\rm OPAL,\, FOPT}), \nn \\
 &\alpha_s(m_\tau^2) = 0.347 \pm 0.025 \qquad ({\rm OPAL,\, CIPT}). 
   \end{align}
 The errors that we quote are dominated by statistics, but they
 include an estimate of the error due to variations of $s_{\rm min}$
  and due to the truncation of the perturbative series. When
 evolved to $m_Z^2$ these results read (with $N_f=5$)
  \begin{align}
 &\alpha_s(m_Z^2) = 0.1191 \pm 0.0022 \qquad ({\rm OPAL, \, FOPT}), \nn \\
 &\alpha_s(m_Z^2) = 0.1216 \pm 0.0027 \qquad ({\rm OPAL, \, CIPT})  .
 \end{align}

The recently published re-analysis of the ALEPH data\cite{Davieretal14} corrects the
correlation matrices and provides us with reliable spectral functions
that have errors significantly smaller than those of OPAL.  At
present, ALEPH's can be considered as the best data set.

The analysis within our framework  based on the new version of the ALEPH spectral
functions was first presented in
Ref.~\refcite{alphas2014}. The main difference with respect to results
derived from OPAL data is in the uncertainties.  The significantly smaller
uncertainties of ALEPH spectra not only translate into smaller
uncertainties for the parameters of the fit, they also resolve any
possible ambiguity due to multiple minima in the fit quality.  In general,
$\alpha_s$ and the $V$ and $A$ channel DV parameters are much better
determined.  The analysis based on the Markov-chain Monte Carlo as
well as the physical tests of the outcome of the fits indicate that
the data are sufficient to constrain the parameters of the fits with
reasonable accuracy,\footnote{The number of parameters varies from 5
  to 13, depending of the specific fit set-up one considers.} 
contrary to the speculation that  ``one has a too large number of free
parameters to be fitted'' made in Ref.~\refcite{PichAachen}. For the
same reasons, there is nothing that indicates that our uncertainties
are underestimated.

The final value of $\alpha_s$ in the ALEPH-based analysis is obtained
from a fit to the vector channel combining the three weight functions
of Eq.~(\ref{moments}). The choice for this fit is based on the
wider stability range against variations of $s_{\rm min}$, but other results are
compatible within errors.\footnote{We should remark that the fit set-up
discussed in Sec. 7 of Ref.~\refcite{PRS16} does not correspond to the fit that gives
our  $\alpha_s$ value.}  We find for $\alpha_s$ in the \MSbar\ and with $N_f=3$
 \begin{align}
& \alpha_s(m_\tau^2) = 0.296 \pm 0.010 \qquad ({\rm ALEPH,\, FOPT}), \nn \\
 &\alpha_s(m_\tau^2) = 0.310 \pm 0.014 \qquad ({\rm ALEPH,\, CIPT}).  
   \end{align}
As before, errors  are dominated by statistics but include an estimate of the error due to varying the $s_0$ window and the  truncation of the perturbative series. Evolving
these results to $m_Z^2$ we find (with $N_f=5$)
  \begin{align}
 &\alpha_s(m_Z^2) = 0.1155 \pm 0.0014 \qquad ({\rm ALEPH, \, FOPT}), \nn \\
&\alpha_s(m_Z^2) = 0.1174 \pm 0.0019 \qquad ({\rm ALEPH, \, CIPT})  .
 \end{align}

\begin{figure}[ht]
\begin{center}
\subfigure[OPAL-based analysis.]{\includegraphics[width=.6\columnwidth,angle=0]{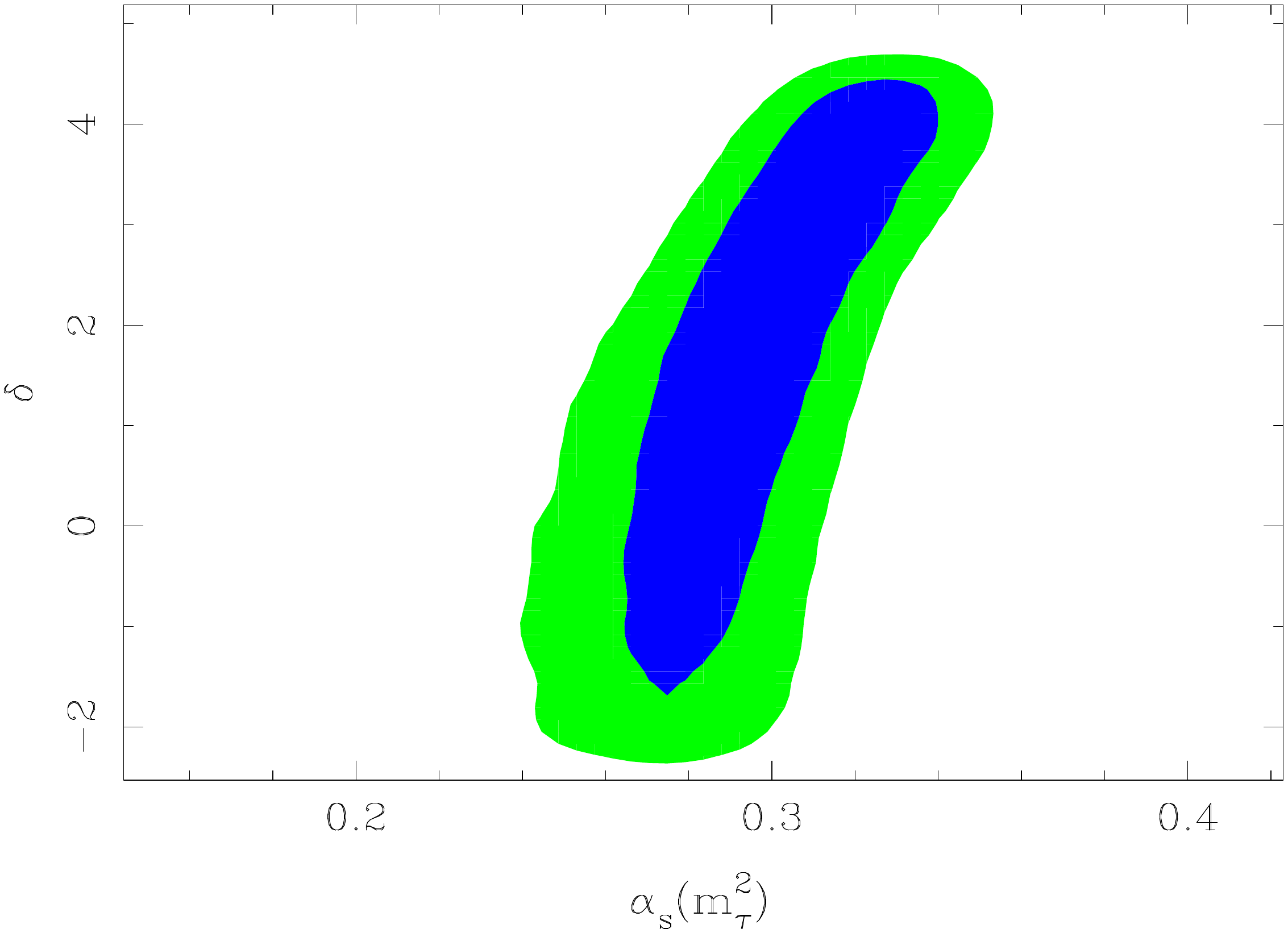}\label{asdvOPAL}}
\subfigure[ALEPH-based analysis.]{\includegraphics[width=.6\columnwidth,angle=0]{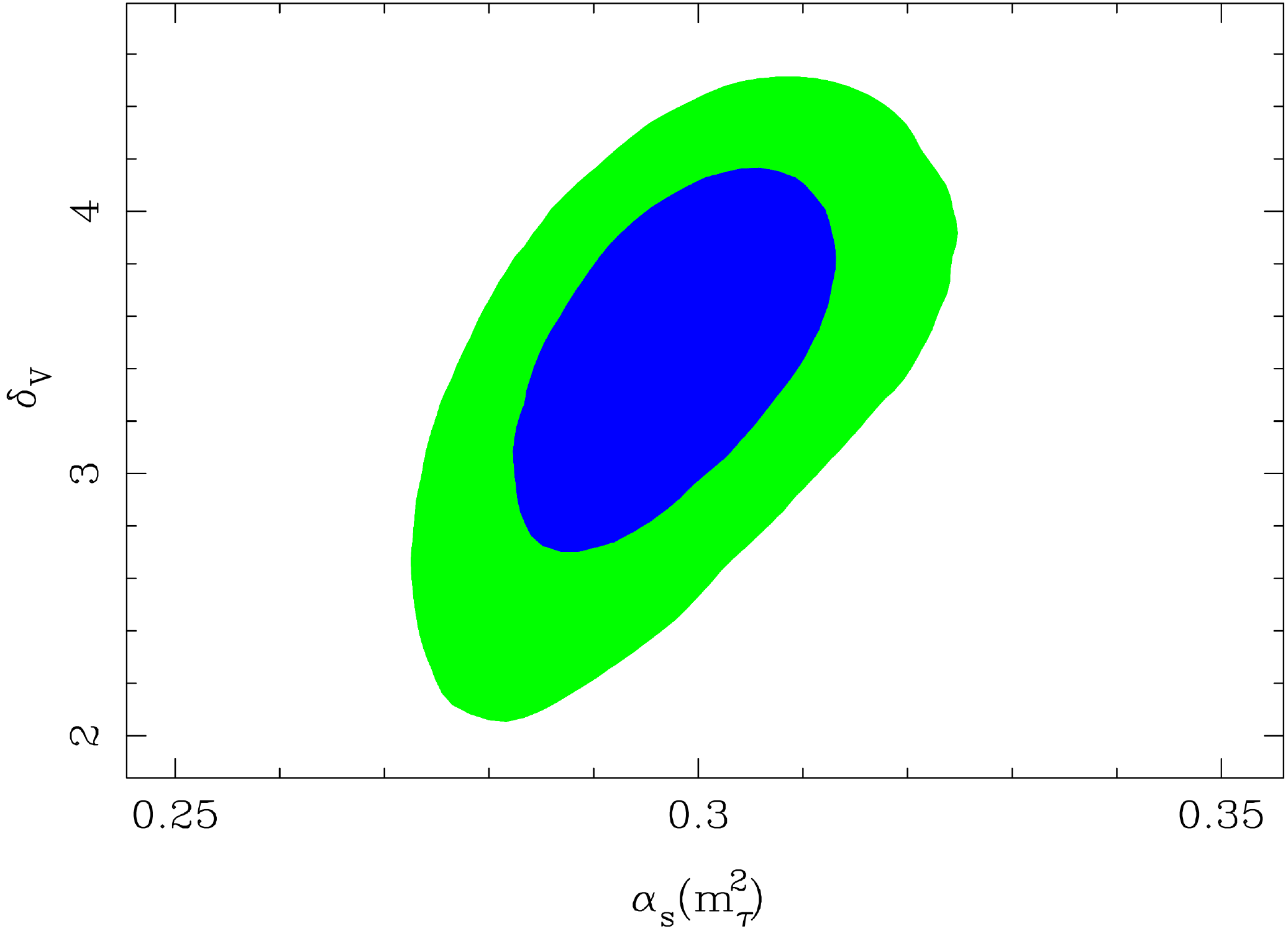}\label{asdvALEPH}}
\vspace*{8pt}
\caption{Two dimensional contour plots for 68\% and 95\% confidence levels in the $\alpha_s$--$\delta_V$ plane. On the left are the OPAL-based results whereas the right-hand panel shows ALEPH-based ones. Note the different  scales in the two plots.}
{\protect\label{ConfFig}}
\end{center}     
\end{figure}

Uncertainties in $\alpha_s$ are smaller when using the ALEPH
data. However, the improvements in other parameters such as, for
example, $\delta_V$ of Eq.~(\ref{DVpar}), are more significant. As an
illustration, Fig.~\ref{ConfFig} shows the allowed intervals for
$\alpha_s$ and $\delta_V$ within 68\% and 95\% confidence levels from
OPAL- and ALEPH-based analyses. The improvement in the vertical spread
of Fig.~\ref{asdvALEPH} is  impressive (note the different scales).

\subsection{Final values}

\begin{figure}[ht]
\begin{center}
\subfigure[$\alpha_s(m_\tau^2)$ FOPT values.]{\includegraphics[width=.5\columnwidth,angle=0]{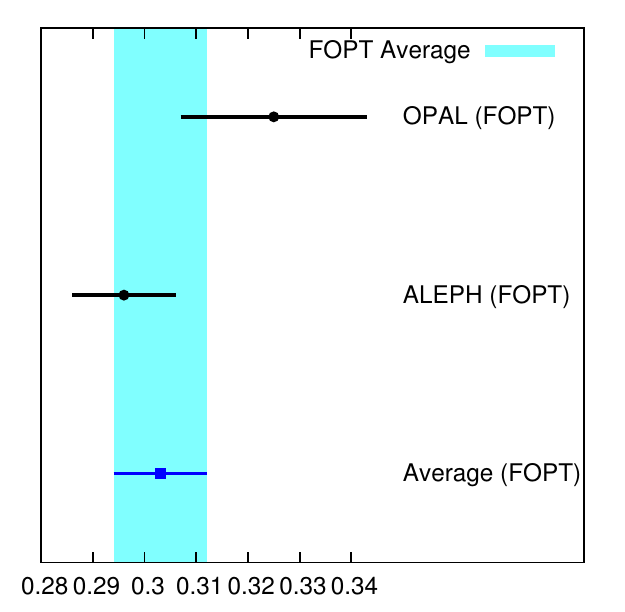}}
\subfigure[$\alpha_s(m_\tau^2)$ CIPT values. Also shown is the FOPT average.]{\includegraphics[width=.475\columnwidth,angle=0]{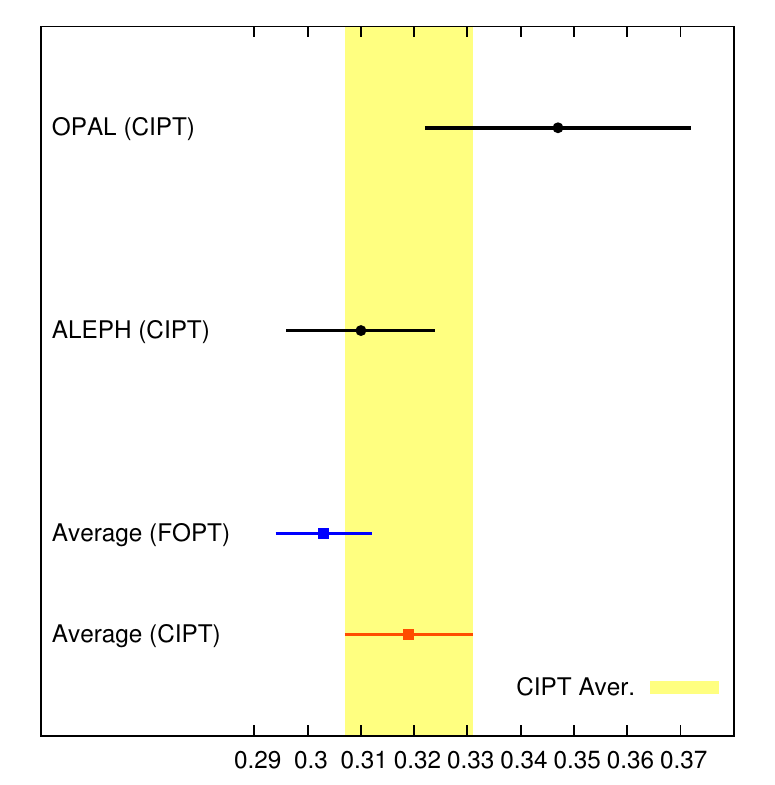}\label{AveragesCIPT}}
\vspace*{8pt}
\caption{Comparison between the $\alpha_s(m_\tau^2)$ values obtained from the OPAL-\protect\cite{alphas2012} and ALEPH-based\protect\cite{alphas2014} analysis.  FOPT results are shown in the left-hand panel, CIPT results in the right-hand panel. Weighted averages given in Eq.~(\ref{alphasaverages}) are also shown for comparison.   }
{\protect\label{alphasComparison}}
\end{center}     
\end{figure}

The results for $\alpha_s$ obtained from ALEPH data tend to be lower than those from
the analysis of OPAL data.  However, within uncertainties, they are compatible. 
Since the spectral functions  are virtually uncorrelated it is legitimate to 
perform a weighted average of our results. The weighted average, which 
we consider our final result for $\alpha_s$, gives (\MSbar\, $N_f=3$)
\begin{align}
 &\alpha_s(m_\tau^2) = 0.303 \pm 0.009 \,\,\,\, (\mbox{ALEPH and OPAL, FOPT}), \nn \\
& \alpha_s(m_\tau^2) = 0.319 \pm 0.012 \,\,\,\, (\mbox{ALEPH and OPAL, CIPT}), \label{alphasaverages} 
 \end{align}
 and at the $Z$ boson mass scale (\MSbar, $N_f=5$)
\begin{align}
 &\alpha_s(m_Z^2) = 0.1165 \pm 0.0012 \,\,\,\, (\mbox{ALEPH and OPAL, FOPT}), \nn \\
& \alpha_s(m_Z^2) = 0.1185 \pm 0.0015 \,\,\,\, (\mbox{ALEPH and OPAL, CIPT}).  \label{alphasaverages2} 
 \end{align}
A visual account of the individual results at the $\tau$ mass scale as
well as the FOPT and CIPT averaged values is given in
Fig.~\ref{alphasComparison}. The averages are compared in
Fig.~\ref{AveragesCIPT}. This comparison shows an overall consistent 
picture, with results compatible with each other and with the average.
The residual difference between CIPT and FOPT remains at the same order 
of the individual uncertainties.

\subsection{A (failed) attempt to fit the spectral functions}

A criticism that has been raised against our analysis strategy regards
the use of moments of $w_1=1$ integrated up to several different $s_0$
values inside the window [$s_{\rm min},s_{\rm max}$]. Besides the
integral of the spectral function up to a certain $s_0$, such a fit
includes information about the shape of the spectral functions which,
of course, plays a role in the extraction of the DV
parameters. Clearly, because we use a FESR of the type shown in
Eq.~(\ref{FESR}), the experimental data below $s_{\rm min}$ also enter
the fit.  In view of this fact, it may be legitimate to question
whether the integral over the data is putting constraints on our value
of $\alpha_s$ and to what extent these constraints come from the shape
of the spectral function itself. In fact, it is the integrated
spectral function that makes a crucial difference in the $\alpha_s$
extraction. For this reason, equating our procedure to a mere fit of
the spectral functions would be very misleading.\cite{PichWSCERN}

An exercise that can be helpful in understanding what is constraining
 the $\alpha_s$ values that we obtain is  to perform a direct
 fit to the actual spectral functions in the interval $s_{\rm
     min}<s<s_{\rm max}$. This fit excludes all experimental
 information for $s<s_{\rm min}$. Obviously, we do not advocate the
 use of such a fit in an $\alpha_s$ extraction, since one would be
 ignoring an important part of the spectral functions --- a part which plays
a key role in obtaining the total weighted spectral integral.
 However, the
 inclusion of DVs in our description of the QCD correlators allows, at
 least in principle, for a fit of this type. A direct fit to the ALEPH
 vector spectral function in the window $1.575~{\rm GeV}^2 \leq s
 \leq m_\tau^2$ produces an acceptable fit ($\chi^2/dof= 1.62$), but
 this fit can barely put any constraint on the strong coupling: it
 gives $\alpha_s(m_\tau^2) = 0.3\pm 0.1$ (for FOPT).  Moreover, the
 results are essentially meaningless since the correlations between
 the fit parameters are huge. For instance, $\alpha_s$ is 96\%
 correlated with $\delta_V$.\footnote{In all $\alpha_s$ analysis from
   $\tau$ decays (including ours) the value of $\alpha_s$ turns out to
   be correlated with non-perturbative parameters. However, these
   correlations reach at most $\sim 65$\%.} We also observe
a rather poor agreement between theory and experiment in FESR obtained
from the results of this fit.

What can be learnt from this exercise is that the experimental
information from regions below $s_{\rm min}$, that enter our fits
through the sum rule of Eq.~(\ref{FESR}), is absolutely crucial to our
$\alpha_s$ determination. Given the present statistical errors, the
shape of the spectral function only very weakly constrains the value
of $\alpha_s$. This happens because the spectral function itself in
the larger $s$ region has a weak dependence on $\alpha_s$ and, hence,
more strongly constrains the DV parameters.  Therefore, the dominant
source of constraint on $\alpha_s$ is the one imposed, through
analyticity, by the large contribution in the weighted integral from
the low-$s$ region.  The inclusion of the integral over the data
reduces the uncertainty by an order of magnitude and leads to a good
match in the FESR.  In conclusion, what constrains the value of
$\alpha_s$ is the integral over the data --- as it should be --- and
not the shape of the spectral functions.

\section{Preliminary combined analysis}
\label{Comparison}

It is reasonable to assume that the spectral functions obtained by OPAL and ALEPH are uncorrelated. Given the fact that the outcome of our analysis
is fully compatible in the two cases, the averages of
Eqs.~(\ref{alphasaverages}) and~(\ref{alphasaverages2}) are justified,
as a first approximation. However, they do not guarantee that a set of
parameters that gives a good description of both data sets exist. The
most rigorous way to combine all the experimental information is to
fit to both data sets simultaneously.  Although it is expected that
the results will be closer to ALEPH ones, given the smaller
uncertainties, it remains a non-trivial test to check that a single
parameter set can describe both data sets reasonably well.

Here we briefly discuss   preliminary results of such a combined
analysis. In this first exploration we fit only to the $V$
channel, with moments of $w_1=1$. The $\chi^2$ to be minimised
is simply given by
\beq
\chi^2 = \chi^2_{\rm ALEPH} + \chi^2_{\rm OPAL},
\eeq
since the correlations between different data sets are assumed negligible.
In this type of fit, one has, in principle, two different $s_0$ windows 
 to consider: one for the OPAL data and one for ALEPH's.
Fit results should be independent of variations in any of the fit windows.

\begin{figure}[!ht]
\begin{center}
\includegraphics[width=.5\columnwidth,angle=0]{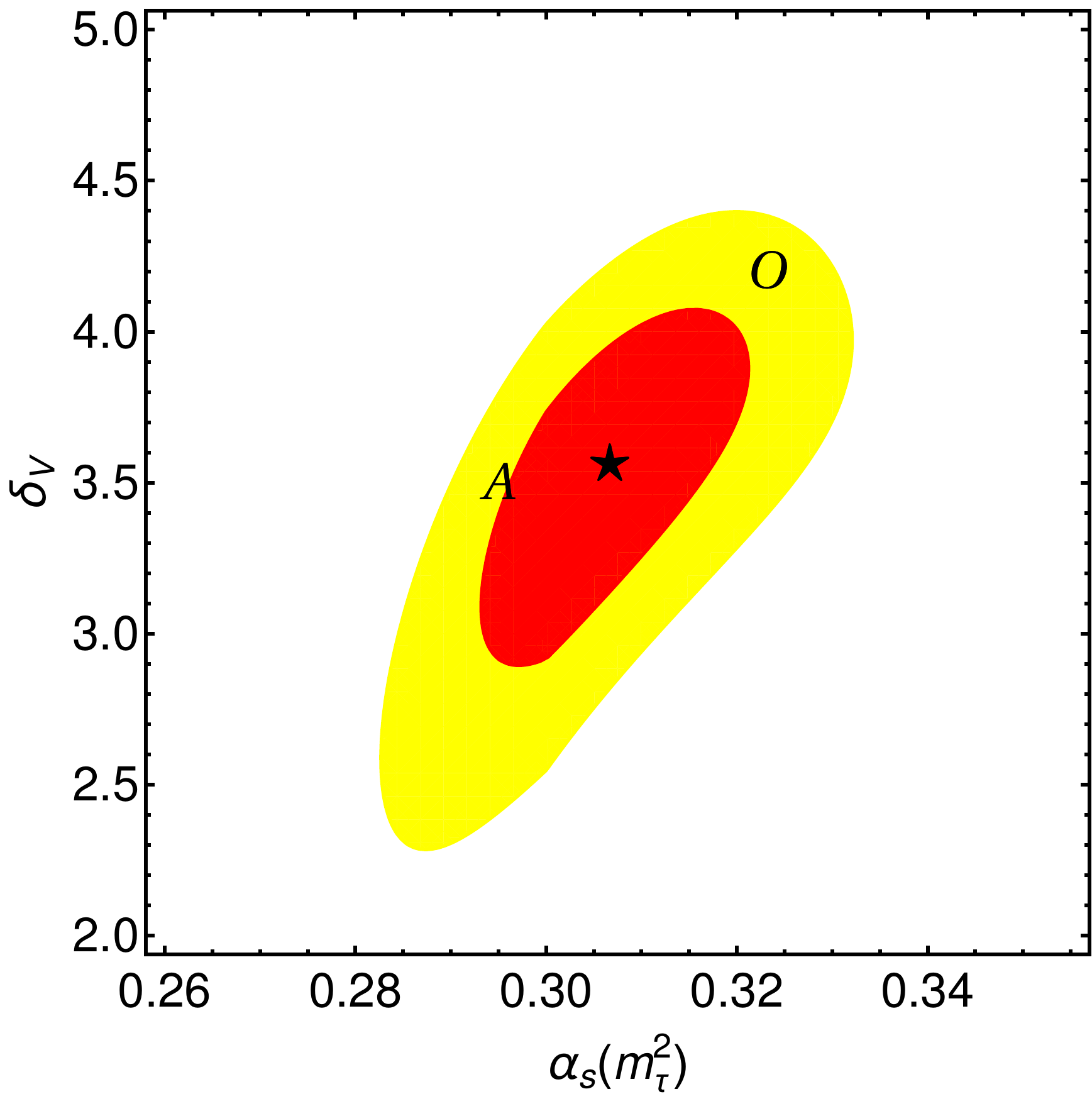}
\vspace*{8pt}
\caption{Two-dimensional contour plots for $68$\% and $95$\% confidence levels in the   $\alpha_s(m_\tau^2)$--$\delta_V$ plane for the combined fit shown in the second row of Table \protect\ref{tab1}. The star marks the central values of the fit, $A$ gives the values of a fit to ALEPH data whereas $O$ gives the central values corresponding to a fit to OPAL data.}
{\protect\label{CombinedFit}}
\end{center}     
\end{figure}

Results for three different fits, performed in different $s_0$ windows, 
are shown in Table~\ref{tab1}. A representative two-dimensional contour plot
on the $\delta_V$--$\alpha_s(m_\tau^2)$ plane is given in Figure~\ref{CombinedFit}.
The results show three main features:
\begin{itemlist}
 \item It is possible to find parameter values that give a good description of the two data sets simultaneously. The $\chi^2$ values obtained are acceptable (if a bit too small) and $p$ values are close to unity.
 \item A comparison with the results of Refs.~\refcite{alphas2012} and~\refcite{alphas2014} shows that $\alpha_s$ values are more stable in this combined analysis than in fits to a single data set.
\item $\alpha_s$ values tend to be slightly larger than the weighted average, but there is good agreement within errors.
\item Values of DV parameters, on the hand, are very close to values obtained in the analysis of ALEPH data.
\end{itemlist}

\begin{table}[!ht]
\tbl{Preliminary results for combined fits to ALEPH and OPAL data with
  $w(x)=1$, vector channel only, FOPT. The ALEPH and OPAL data are assumed to be
  uncorrelated. All correlations inside the individual data sets are taken into account. The values $s_{\rm min}^{\rm AL}$ and $s_{\rm min}^{\rm OP}$   show the choice for
  the minimum $s_0$ in the ALEPH and OPAL data respectively. Uncertainties are solely statistical.}
{\begin{tabular}{@{}cccccccc@{}} 
\toprule
$s_{\rm min}^{\rm OP}$ [GeV$^2$] &$s_{\rm min}^{\rm AL}$ [GeV$^2$] & $\chi^2$/dof &$\alpha_s(m_\tau)$ & $\delta_V$ & $\gamma_V$ &$\alpha_V$ & $\beta_V$  \\
\colrule
1.5 &1.50 & $46.4/70$ & 0.308(10) & 3.43(37) & 0.63(23) & $-0.98(66)$  & 3.60(34)\\
1.5 &1.55 & $43.9/68$ & 0.307(09) & 3.56(38) & 0.57(23) & $-1.19(64)$  & 3.71(33)\\
1.6 &1.60 & $41.0/63$ & 0.308(10) & 3.50(40) & 0.60(23) & $-1.10(85)$  & 3.66(43)\\
\botrule
\end{tabular}\label{tab1} }
\end{table}

\section{Conclusions}
\label{Conclusions}

We have discussed a new strategy for the QCD analysis of hadronic
$\tau$ decay data.  The main advantage of this strategy is that it
allows for a fully self-consistent analysis, without relying on any
untested assumption. This can be achieved thanks to the introduction
of a physically motivated parametrisation for the DV contributions. Our analysis
does not rely only on pinched moments, avoiding
contamination by higher-order OPE condensates.

The strategy was applied to OPAL data at first\cite{alphas2012} and, more recently, to
ALEPH data as well.\cite{alphas2014} In both cases one can extract $\alpha_s$ together
with non-perturbative contributions such as the dimension 6 and 8 OPE
condensates and DV parameters.  The results from the two data sets are
compatible. Since the data are uncorrelated we can perform weighted
averages and, at present, our recommended value of $\alpha_s$ is
obtained as an  average from the ALEPH- and OPAL-based determinations.
These values can be found in Eqs.~(\ref{alphasaverages}) and~(\ref{alphasaverages2}).

We have performed extensive statistical and physical tests on the
results of our fits.  They indicate that the results are reliable and
satisfy various self-consistency checks. In addition, the
uncertainties obtained are realistic and  the number of
parameters that we fit is manageable.  Among the self-consistency checks
passed by our results are those given by the  Weinberg sum
rules.\cite{alphas2011,alphas2012,alphas2014,PerisMainz}

Finally, we have performed preliminary simultaneous fits to ALEPH and
OPAL data.  The results obtained are encouraging because they are
stable and give a good representation of both data sets. This lends
support to the validity of the theoretical description and to the
compatibility of the two data sets.

Further progress would require a better theoretical understanding of
DVs and/or better spectral functions. In principle, with a dedicated
effort, the latter could be extracted from Belle and BaBar data and
could lead to significant progress in the field.

\section*{Acknowledgments}
It is our pleasure to thank the Mainz Institute for Theoretical
Physics, the Joahannes Gutenberg Universit\"at, and, in particular,
the organisers of this very fruitful workshop.  DB's work is supported
by the S\~ao Paulo Research Foundation (FAPESP) grant 15/20689-9 and by CNPq (grant 305431/2015-3). DB's
attendance to the workshop was partially funded by the Alexander von
Humboldt Stiftung/Foundation. MG is supported in part by the
U.S. Department of Energy.
SP is supported by CICYTFEDER-FPA2014-55613-P, 2014-SGR-1450. 
 KM is supported by a grant from the Natural
Sciences and Engineering Research Council of Canada.

\end{document}